\documentclass[pra,twocolumn,amsmath,amssymb,superscriptaddress]{revtex4-1}
\usepackage{graphicx}
\usepackage{color}
\usepackage{braket}
\usepackage{amsmath}
\usepackage{amsfonts}
\usepackage{natbib}
\usepackage{bbold}
\usepackage{bm}
\usepackage{verbatim}

\usepackage{amsmath,amssymb,mathrsfs}
\usepackage{bbm}
\usepackage[dvipsnames]{xcolor}

\usepackage[justification=Justified]{caption}

\begin{document}

\title{Majorana zero modes in gate-defined germanium hole nanowires}

\author{Katharina Laubscher}

\author{Jay D. Sau}

\author{Sankar Das Sarma}

\affiliation{Condensed Matter Theory Center and Joint Quantum Institute, Department of Physics, University of Maryland, College Park, MD 20742, USA}

\date{\today}

\begin{abstract}
We theoretically study gate-defined one-dimensional channels in planar Ge hole gases as a potential platform for non-Abelian Majorana zero modes. We model the valence band holes in the Ge channel by adding appropriate confinement potentials to the 3D Luttinger-Kohn Hamiltonian, additionally taking into account a magnetic field applied parallel to the channel, an out-of-plane electric field, as well as the effect of compressive strain in the parent quantum well. Assuming that the Ge channel is proximitized by an $s$-wave superconductor (such as, e.g., Al) we calculate the topological phase diagrams for different channel geometries, showing that sufficiently narrow Ge hole channels can indeed enter a topological superconducting phase with Majorana zero modes at the channel ends. We estimate the size of the topological gap and its dependence on various system parameters such as channel width, strain, and the applied out-of-plane electric field, allowing us to critically discuss under which conditions Ge hole channels may manifest Majorana zero modes. Since ultra-clean Ge quantum wells with hole mobilities exceeding one million and mean-free paths on the order of many microns already exist, gate-defined Ge hole channels may be able to overcome some of the problems caused by the presence of substantial disorder in more conventional Majorana platforms.
\end{abstract}

\maketitle

\section{Introduction}

Germanium is emerging as a promising material platform for various quantum-technological applications~\cite{Scappucci2021}. In particular, Ge hole spin qubits are prominent candidates for spin-based quantum information processing due to favorable properties such as weak hyperfine interaction, large and tunable spin-orbit energies that enable fast qubit operations, and tunable effective $g$ factors~\cite{Hu2007,Hu2012,Ares2013,Watzinger2018,Li2018,Hendrickx2020,Jirovec2021,Froning2021,Wang2022,Hofmann2019}. While many early studies focused on hole spin qubits defined in Ge/Si core/shell or Ge hut nanowires, substantial experimental progress has recently established Ge two-dimensional hole gases (2DHGs) as an extremely clean and versatile platform for gate-defined hole spin qubits. Indeed, ultra-high quality Ge 2DHGs with hole mobilities exceeding one million and mean-free paths on the order of tens of microns have been reported~\cite{Sammak2019,Lodari2022,Myronov2023,Stehouwer2023}. The two-dimensional geometry of the parent quantum well additionally facilitates scalability, with important recent experiments realizing gate-defined quantum dot arrays~\cite{Lawrie2020,Riggelen2020} and multi-qubit logical operations~\cite{Hendrickx2020b,Hendrickx2021}.

Going beyond standard spin-qubit applications, recent experiments report the fabrication of Ge-based semiconductor/superconductor hybrid devices~\cite{Vries2018,Hendrickx2018,Hendrickx2019,Vigneau2019,Aggarwal2021} with hard proximity-induced superconducting gaps~\cite{Tosato2022}. Such devices hold significant potential for the realization of Majorana zero modes (MZMs) due to the extremely high quality of the underlying Ge. Indeed, in standard semiconductor/superconductor hybrid devices based on InAs or InSb~\cite{Mourik2012,Rohkinson2012,Das2012,Deng2012,Lee2012,Churchill2013,Deng2016,Deng2018}, the presence of substantial disorder has hampered any conclusive observation of MZMs so far~\cite{Sau2012,Sau2013,Chiu2017,Pan2020,Pan2020b,Dassarma2021,Ahn2021,Dassarma2023,Dassarma2023b,Dassarma2023c}, although a very recent Microsoft experiment~\cite{Aghaee2022} reports the observation of small topological gaps in very limited regions of the parameter space (of gate voltage and magnetic field) in InAs/Al devices. In contrast, a recent experiment~\cite{Stehouwer2023} shows that the mobility in Ge 2DHGs can be 50-100 times larger than the electron mobility in InAs. This makes Ge nanowires a plausible candidate platform for topological MZMs since the Ge system is already ultra-clean.

With this motivation, we investigate the prospects for the realization of Ge-based MZMs and study one-dimensional (1D) Ge hole channels obtained by electrostatic confinement of a Ge 2DHG (see Fig.~\ref{fig:setup} for a schematic illustration) as a potential platform for MZMs. Importantly, since the valence band holes in Ge effectively carry spin $3/2$, Ge hole nanowires show qualitative and quantitative differences compared with standard electron nanowires (e.g., InAs or InSb). For example, the spin-orbit interaction (SOI) in Ge hole nanowires is predicted to reach values on the order of meV~\cite{Kloeffel2011,Kloeffel2018}, which is much larger than what is expected in, e.g., InAs. This, in principle, enhances the topological gap, other things being equal. Additionally, both the SOI as well as the effective $g$ factor in hole nanowires exhibit a strong dependence on local details such as the wire geometry, leading to an overall richer behavior of hole nanowires compared to electron nanowires.

To model the valence band holes in the Ge channel, we start from the standard 3D Luttinger-Kohn Hamiltonian, to which we add appropriate confinement potentials. We account for a magnetic field applied parallel to the channel, an out-of-plane electric field, the effect of compressive strain in the parent quantum well, and proximity-induced superconductivity due to the presence of a thin Al strip in the vicinity of the channel. By numerically calculating the associated topological phase diagrams through the exact solutions of the appropriate Bogoliubov-de Gennes (BdG) equations, we show that sufficiently narrow Ge hole channels can indeed enter a topological superconducting phase with MZMs at the ends of the channel. We present results for various wire geometries, for different values of strain, and for different strengths of the external electric field. We estimate the maximal topological gaps to be on the order of tens of $\mu$eV in narrow channels, which is comparable to the predicted as well as recently reported topological gaps in InAs nanowires~\cite{Aghaee2022}. As a general trend, we find that the topological gaps grow as the channel width decreases and as strain is reduced. In wide channels, the main limiting factor is the small effective in-plane $g$ factor of the lowest-energy Ge subband, which pushes the topological phase transition to large magnetic fields. For all of the considered wire geometries, we find that the external electric field provides an additional tuning knob that can be adjusted in order to maximize the topological gap for a given geometry.

We note that previous theoretical works have explored Ge/Si core/shell nanowires~\cite{Maier2014} and planar Josephson junctions based on Ge 2DHGs~\cite{Luethi2022,Luethi2023} as potential platforms for MZMs. Furthermore, MZMs in hole nanowires based on materials other than Ge have been theoretically studied in Refs.~\cite{Mao2012,Sau2012,Liang2017}. However, we believe that the setup described in the present work is ideally suited to take maximal advantage of the already existing ultra-clean planar Ge quantum wells. Furthermore, electrostatically defined Ge hole channels of high quality have already been realized in proof-of-principle experiments~\cite{Mizokuchi2018}, putting this setup well within experimental reach. 

The rest of this paper is organized as follows. In Sec. II, we describe the basic model for a gate-defined Ge hole channel that we use in our numerical simulations. In Sec.~III, we describe our calculations and present our numerical results. Finally, we conclude in Sec.~IV.

\section{Model}

The valence band holes of 3D bulk Ge are well described by the isotropic Luttinger-Kohn (LK) Hamiltonian ~\cite{Luttinger1956,Winkler2003}
\begin{equation}
\mathcal{H}_{LK}=\frac{\hbar^2}{m}\left[
\left(\gamma_1 + \frac{5 \gamma_s}{2}\right)\frac{\bm{k}^{2}}{2}
- \gamma_s \left( \bm{k} \cdot \bm{J} \right)^2 
\right]-\mu, 
\label{eq:lk_hamiltonian_isotropic}
\end{equation}
where $m$ is the bare electron mass, $\gamma_1=13.35$, $\gamma_2=4.25$, and $\gamma_3=5.69$ are the Luttinger parameters for Ge, $\gamma_s=(\gamma_2+\gamma_3)/2$, $\bm{k}=(k_x,k_y,k_z)$ is the vector of momentum, $\bm{J}=(J_x,J_y,J_z)$ is the vector of spin-$3/2$ operators, and $\mu$ is the chemical potential. Since $\gamma_3-\gamma_2\ll \gamma_1$ in Ge, anisotropic corrections to the LK Hamiltonian are small and the isotropic approximation given in Eq.~(\ref{eq:lk_hamiltonian_isotropic}) is well justified. We note that here and throughout this paper, since we are considering hole excitations instead of electrons, we omit a global minus sign in front of all Hamiltonian terms for convenience.

\begin{figure}[tb]
	\centering
	\includegraphics[width=\columnwidth]{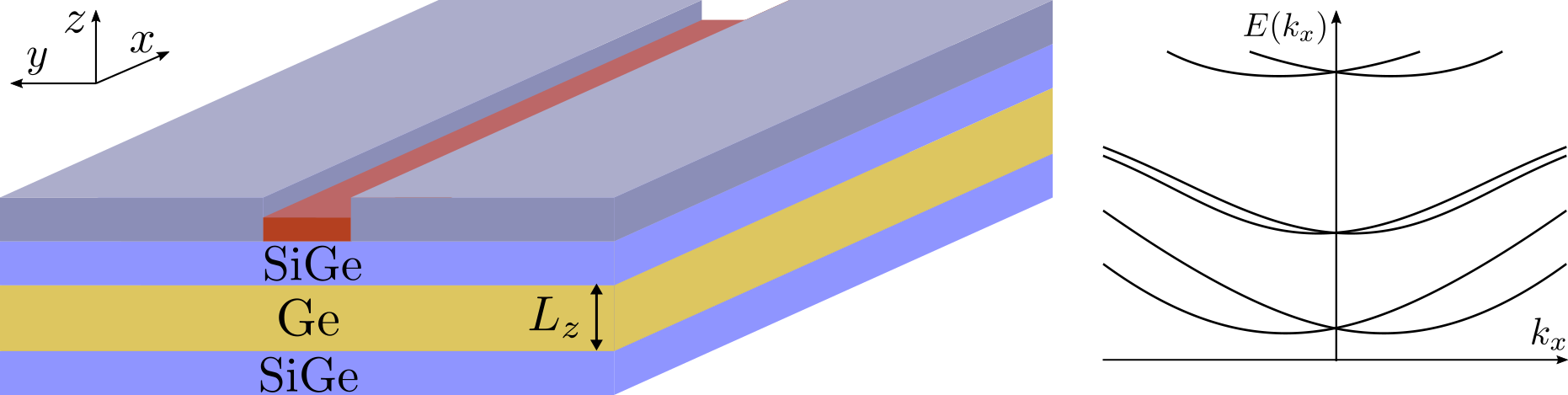}
	\caption{Left: Sketch of a Ge quantum well of thickness $L_z$ (yellow) sandwiched between two layers of SiGe (blue). The 2DHG in the quantum well is further confined into a quasi-1D geometry by electrostatic gates (gray). If the channel is proximitized by a superconductor (red) and a magnetic field is applied along the direction of the channel, MZMs can emerge at the channel ends. Right: Schematic low-energy spectrum of a 1D Ge hole channel at zero magnetic field and in the absence of a superconductor (note that a global minus sign was omitted from the hole spectrum).}
	\label{fig:setup}
\end{figure}

In the following, we consider a 2D Ge quantum well of thickness $L_z$ encapsulated between two layers of Si$_{1-x}$Ge$_x$, see Fig.~\ref{fig:setup}. We model the confining potential arising at the interface between the Ge and the SiGe by an infinite hard-wall potential along the $z$ direction:
\begin{equation}
\mathcal{H}_{\mathrm{conf},\perp}(z)=\begin{cases} 0 & 0<z<L_z,\\\infty & \mathrm{otherwise.}\end{cases}
\end{equation}
At zero in-plane momentum, the confinement to two dimensions leads to an energy splitting between bands with spin projection $\pm 3/2$ along the $z$ direction (heavy holes, HHs) and bands with spin projection $\pm 1/2$ along the $z$ direction (light holes, LHs), with the energy of the latter becoming higher due to confinement. For finite in-plane momentum, the LK Hamiltonian mixes HHs and LHs, but, at low energies, the lowest subband retains predominantly HH character~\cite{Winkler2003}.

In typical Ge/SiGe quantum wells, the Ge is compressively strained due to the lattice mismatch between the Ge and the Si$_{1-x}$Ge$_x$. The strain is modeled by the Bir-Pikus (BP) Hamiltonian~\cite{Bir1974}
\begin{equation}
\mathcal{H}_{BP}=-E_s J_z^2,\label{eq:strain}
\end{equation}
where the strain energy $E_s>0$ increases with the percentage $1-x$ of Si in the Si$_{1-x}$Ge$_x$. For typical values $x\in(0.6,0.9)$, $E_s$ is of the order of tens of meV~\cite{Lodari2022}. Furthermore, we consider an external electric field of strength $\mathcal{E}_z$ that is applied along the $z$ direction (i.e., out-of-plane),
\begin{equation}
\mathcal{H}_{el}=-e \mathcal{E}_z z,
\end{equation}
where $e$ is the positive elementary charge. The electric field breaks inversion symmetry and leads to spin-orbit interaction (SOI) of Rashba type, which has a cubic dependence on the in-plane momentum in planar Ge~\cite{Winkler2000,Winkler2008,Moriya2014,Marcellina2017,Terrazos2021,Mizokuchi2017,Chou2018} that becomes linear upon further confinement to 1D~\cite{Kloeffel2011,Kloeffel2018,Hao2010,Gao2020,Froning2021b,Adelsberger2021,Adelsberger2022}, see also below. Additionally, the electric field tends to push the low-energy hole wave functions towards the top of the quantum well, which introduces an additional length scale $l_\mathcal{E}=(\hbar^2\gamma_1/2me\mathcal{E}_z)^{1/3}$ into the problem.

To obtain a quasi-1D geometry, the 2DHG is further confined by electrostatic gates from the sides. Since the precise form of the smooth confinement potential is neither known exactly nor trivial to model numerically, we restrict ourselves to discussing the two extreme cases of (1) infinite hard-wall confinement and (2) parabolic confinement along the $y$ direction. In case (1), the confinement potential takes the form
\begin{equation}
\mathcal{H}_{\mathrm{conf},\parallel}^{(1)}(y)=\begin{cases} 0 & 0<y<L_y,\\\infty & \mathrm{otherwise,}\end{cases}
\end{equation}
where $L_y$ is the width of the channel. In case (2), the confinement potential is taken to be
\begin{equation}
\mathcal{H}_{\mathrm{conf},\parallel}^{(2)}(y)=\frac{\hbar^2\gamma_1}{2ml_y^4}y^2,
\end{equation}
where $l_y$ is the harmonic confinement length. In this case, the `width' of the channel is not a well-defined quantity. Whenever we compare between the two different confinement potentials, we therefore take the `width' of the parabolic well to be $2l_y$, which corresponds to the width felt by the lowest subband of the parabolic well. In wide channels, where the confinement along the $y$ direction is much weaker than the confinement along the $z$ direction, the system is not far from the 2D limit and the lowest subband has predominantly HH character. However, in narrow channels with two axes of comparably strong confinement, the situation is drastically different as HHs and LHs are strongly mixed even at low energies~\cite{Sercel1990,Csontos2009}. As we will discuss below, this has important implications for the topological phase diagram of the channel. We mention that our confinement models defined by Eqs.~(2), (5), and (6) are simple, but not unreasonable, and enable the notion of a wire width as the controlling parameter for the discussions of our theoretical results. If and when MZM experiments are performed in Ge nanowires, it should be possible to generalize our confinement models to more realistic situations as relevant for the specific experimental samples.

We additionally account for an external magnetic field of strength $B$ along the $x$ direction, i.e., parallel to the 1D channel. The orbital effects associated with the magnetic field lead to an additional term in the bulk LK Hamiltonian,
\begin{align}
&\mathcal{H}_\mathrm{orb}=\frac{\hbar e}{2m}[(\gamma_1+\frac{5\gamma_s}{2})(\frac{e}{h}\bm{A}^2+2\bm{k}\cdot\bm{A})-\frac{2\gamma_s e}{\hbar}(\bm{A}\cdot\bm{J})^2\nonumber\\& -4\gamma_s(k_xA_xJ_x^2+(\{k_x,A_y\}+\{k_y,A_x\})\{J_x,J_y\}+\mathrm{c.p.})],\label{eq:orb}
\end{align}
where $\bm{A}$ is the vector potential satisfying $\bm{B}=\bm{\nabla}\times\bm{A}$, $\{A,B\}=(AB+BA)/2$, and where `c.p.' stands for `cyclic permutations'. For our numerical simulations, we fix the gauge to $\bm{A}=(0,0,By)$. Furthermore, the magnetic field leads to a Zeeman splitting of the form~\cite{Luttinger1956,Winkler2003}
\begin{equation}
\mathcal{H}_Z=2\kappa\mu_B B J_x,
\end{equation}
where $\mu_B$ is the Bohr magneton and $\kappa\approx 3.41$ for Ge~\cite{Lawaetz1971}. The total normal-state Hamiltonian that we consider in the remainder of this paper then takes the form $H_0=\int d\bm{r}\,\psi^\dagger(\bm{r})\mathcal{H}_0(\bm{r})\psi(\bm{r})$ with 
$\psi=(\psi_{3/2}$,$\psi_{1/2},\psi_{-1/2},\psi_{-3/2})^T$ 
and
\begin{equation}
\mathcal{H}_0= \mathcal{H}_{LK}+\mathcal{H}_{\mathrm{conf},\perp}+\mathcal{H}_{BP}+\mathcal{H}_{el}+\mathcal{H}_{\mathrm{conf},\parallel}^{(i)}+\mathcal{H}_\mathrm{orb}+\mathcal{H}_Z,\label{eq:normalH}
\end{equation}
where $i=1$ ($i=2$) corresponds to the case of hard-wall (parabolic) confinement along the channel.

The normal-state Hamiltonian $\mathcal{H}_0$ has been studied in some detail by previous works~\cite{Adelsberger2021,Adelsberger2022} focusing mainly on spin-qubit applications, and it is useful to review some of its properties at this point. Up to a global minus sign that we omit in this work, the low-energy band structure of $\mathcal{H}_0$ resembles the one of electrons in a conventional Rashba nanowire, see Fig.~\ref{fig:setup} for an example. Around $k_x=0$, the lowest-energy subspace of $\mathcal{H}_0$ can be described by a simple effective two-band Hamiltonian of the form~\cite{Adelsberger2022}
\begin{equation}
\mathcal{H}_\mathrm{eff}=\frac{\hbar^2 k_x^2}{2\bar{m}}+\frac{1}{2}\left(g_\mathrm{eff}\mu_BB+\frac{\hbar^2k_x^2}{\bar{m}_s}\right)\sigma_x-\alpha_{so}k_x\sigma_y,\label{eq:eff2band}
\end{equation}
where $\bar{m}$ is the effective mass, $g_\mathrm{eff}$ is the effective $g$ factor, $\alpha_{so}$ is the effective spin-orbit coupling strength, $\bar{m}_s$ is an effective spin-dependent mass, and the Pauli matrices $\sigma_i$ with $i\in\{x,y,z\}$ act in the subspace of the two lowest-energy subbands. However, in contrast to the case of spin-1/2 electrons in semiconductor nanowires, the lowest Ge hole subband has contributions from both states with spin projection $\pm 3/2$ (HHs) and $\pm 1/2$ (LHs), with the relative weight of these two contributions depending sensitively on the wire geometry, on the shape of the confinement potentials, and on strain. As a consequence, the effective parameters entering Eq.~(\ref{eq:eff2band}) also show a strong dependence on all of these factors~\cite{Adelsberger2021,Adelsberger2022}, making it generally necessary to solve the full Hamiltonian $\mathcal{H}_0$ to correctly capture these features. Therefore, while we will frequently refer to the effective Hamiltonian $\mathcal{H}_\mathrm{eff}$ [Eq.~(\ref{eq:eff2band})] for intuition, all numerical calculations presented in this work use the full normal-state Hamiltonian $\mathcal{H}_0$ [Eq.~(\ref{eq:normalH})].

Finally, we include a proximity-induced superconducting pairing, which we take to be of the form
\begin{equation}
H_{sc}=\int d\bm{r} \sum_{s=\frac{1}{2},\frac{3}{2}}\Delta_s\,\psi_s^\dagger(\bm{r}) \psi_{-s}^\dagger(\bm{r})+\mathrm{H.c.},\label{eq:sc_pairing}
\end{equation}
where $\Delta_{1/2}$ ($\Delta_{3/2}$) is the superconducting pairing amplitude for LHs (HHs). In the following, we assume for simplicity that the HH and LH pairing amplitudes are equal in magnitude but of opposite sign, i.e., $\Delta_{3/2}=-\Delta_{1/2}\equiv\Delta$~\cite{note1}. With this choice, the size of the effective superconducting gap that is opened in the lowest confinement-induced Ge hole subband is independent of the wire geometry (see below). Indeed, since the precise microscopic description of the proximity-induced superconducting pairing in Ge/superconductor hybrid structures is not known and, in addition, is likely to depend on the details of a particular sample, we focus on a simple description that keeps the number of unknown parameters to a minimum. Nevertheless, generalizing our analysis to unequal pairing amplitudes $|\Delta_{1/2}|\neq  |\Delta_{3/2}|$ is straightforward; the only relevant effect is that the magnitude of the effective superconducting gap that is opened in the lowest confinement-induced subband gets renormalized.

We model the suppression of the proximity-induced superconducting gap due to the applied magnetic field as
\begin{equation}
\Delta=\Delta_0\sqrt{1-\left(\frac{B}{B_c}\right)^2}\Theta(B_c-|B|),
\end{equation}
where $\Delta_0$ is the proximity-induced superconducting gap at zero magnetic field, $B_c$ is the critical magnetic field of the superconductor, and $\Theta$ is the Heaviside step function ensuring that the superconducting gap is zero for any $|B|\geq B_c$. One can take $B_c$ to be the approximate field value where the bulk gap of the parent superconductor is closed by the applied field as observed experimentally. For concreteness, we focus on a Ge/Al heterostructure in this work, where we take the critical field of the Al strip to be $B_c=3$~T~\cite{Nichele2017,Suominen2017}. We note that a more elaborate treatment of the proximity effect should explicitly include the tunneling between the superconductor and the Ge. Such a description could then also capture the regime of strong coupling~\cite{Sau2010,Stanescu2010,Cole2015,Reeg2018}, where the induced superconducting gap as well as the underlying Ge band structure parameters get renormalized~\cite{Adelsberger2023}. However, for the present purpose, a minimal description of the proximity-induced superconducting pairing as given in Eq.~(\ref{eq:sc_pairing}) is sufficient to capture the important qualitative features of the system in the regime of weak coupling between the superconductor and the Ge. If necessary, the theory can be generalized to include the self-energy effect describing the proximity effect, but such a generalization is unnecessary and not useful at this early stage where the goal is to see if the Ge-based MZM platform is a feasible idea or not.

Extracting the exact low-energy behavior of proximitized Ge hole channels in an experimentally realistic setting is a highly non-trivial problem that we will not attempt to solve in completeness (this should not be done without a detailed knowledge of the actual experimental system being used searching for MZMs in Ge nanowires since all the details would matter at a quantitative, but not a qualitative, level). Instead, in the next section, we present model calculations for various wire geometries and a wide range of additional parameters such as strain and electric field, and based on these results we critically discuss the experimental feasibility of MZMs in Ge hole nanowires. While we do provide quantitative estimates for the maximal topological gaps that are achievable in a given wire geometry, our main focus lies on identifying general trends that describe the qualitative behavior of the system.

\begin{figure*}[tb]
	\centering
	\includegraphics[width=\textwidth]{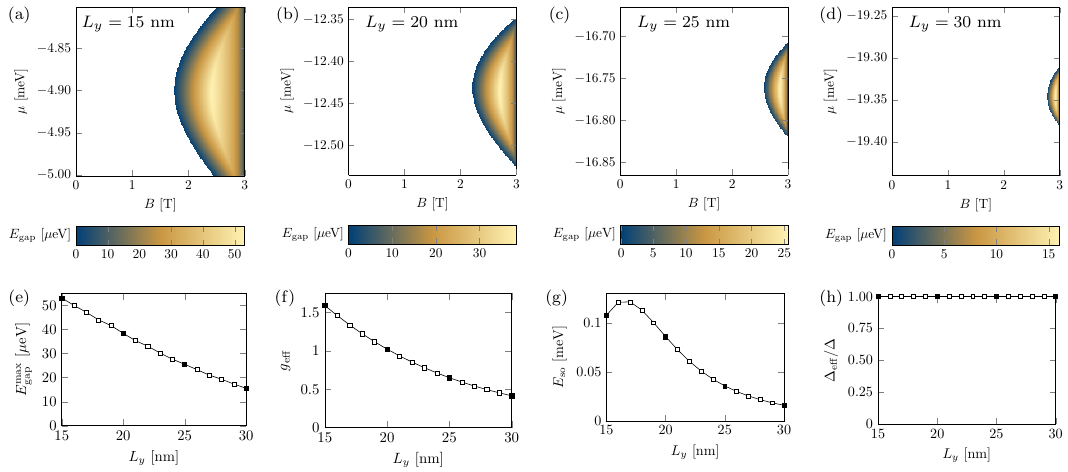}
	\caption{(a-d) Topological phase diagrams obtained by numerically diagonalizing $H=H_0+H_{sc}$ for different widths of the channel $L_y$ (see insets) in the case of hard-wall confinement along the $y$ direction. The white regions correspond to the trivial phase, while the colored regions correspond to the topological phase with the color encoding the size of the bulk gap. (e) Maximal bulk gap in the topological phase in dependence on the channel width. (f-h) Effective $g$ factor at $B=2$~T, effective spin-orbit energy at zero magnetic field, and effective superconducting gap at zero magnetic field in dependence on the channel width. Filled squares correspond to the widths shown in (a-d). The solid lines are a guide to the eye only. We fix $L_z=22$~nm, $\Delta_0=0.1$~meV, $E_s=10$~meV, and $\mathcal{E}_z=0.5$~V$\mu$m$^{-1}$ for all panels.
	}
	\label{fig:phase_diagrams_square}
\end{figure*}

\begin{figure*}[tb]
	\centering
	\includegraphics[width=\textwidth]{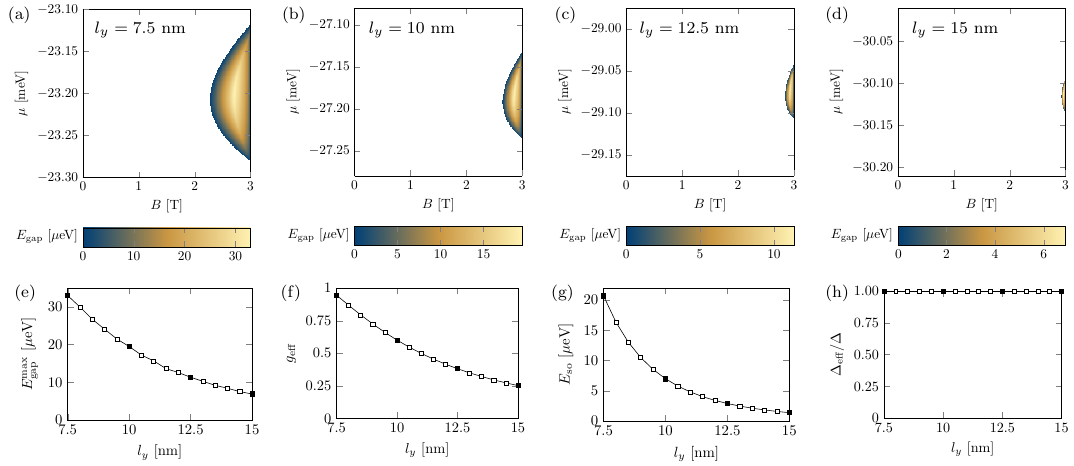}
	\caption{(a-d) Topological phase diagrams obtained by numerically diagonalizing $H=H_0+H_{sc}$ for different confinement lengths $l_y$ (see insets) in the case of parabolic confinement along the $y$ direction. The white regions correspond to the trivial phase, while the colored regions correspond to the topological phase with the color encoding the size of the bulk gap. (e) Maximal bulk gap in the topological phase in dependence on the confinement length. (f-h) Effective $g$ factor at $B=2$~T, effective spin-orbit energy at zero magnetic field, and effective superconducting gap at zero magnetic field in dependence on the confinement length. Filled squares correspond to the widths shown in (a-d). The solid lines are a guide to the eye only. We fix $L_z=22$~nm, $\Delta_0=0.1$~meV, $E_s=10$~meV, and $\mathcal{E}_z=1$~V$\mu$m$^{-1}$ for all panels.}
	\label{fig:phase_diagrams_parabolic}
\end{figure*}

\section{Topological phase diagrams}

The low-energy spectrum of the gate-defined Ge hole channel can be obtained by rewriting the full Hamiltonian $H=H_0+H_{sc}$ in BdG form and expanding its eigenstates in terms of suitable basis functions that solve the confinement problem~\cite{Csontos2009,Kloeffel2011,Kloeffel2018,Adelsberger2021,Adelsberger2022,Milivojevic2021}. Assuming translational invariance along the channel, we write the spatially varying part of these basis functions as $\varphi_{k_x,p,q}(x,y,z)=e^{ik_xx}\varphi_{p}(y)\varphi_{q}(z)$, where
\begin{equation}
\varphi_{q}(z)=\begin{cases} \sqrt{2/L_z}\sin\left(q\pi z/L_z\right)&z\in(0,L_z)\\0&\mathrm{ otherwise }\end{cases}
\end{equation}
with $q\in\{1,2,...\}$ are the eigenfunctions of the infinite square well along the $z$ direction.
Similarly, along the $y$ direction, we use
\begin{equation}
\varphi_{p}^{(1)}(y)=\begin{cases} \sqrt{2/L_y}\sin\left(p\pi y/L_y\right)&y\in(0,L_y)\\0&\mathrm{ otherwise }\end{cases}
\end{equation}
with $p\in\{1,2,...\}$ for the case of hard-wall confinement and 
\begin{equation}
\varphi_{p}^{(2)}(y)=e^{-y^2/2l_y^2}H_{p}(y/l_y)/\sqrt{2^p\sqrt{\pi}l_y p!}
\end{equation}
with $p\in\{0,1,...\}$ for the case of parabolic confinement (here $H_p$ are the Hermite polynomials). For our numerical simulations, we project the full BdG Hamiltonian into the subspace spanned by the first 10 basis functions for each spatial direction, which results in an $800\times800$ effective Hamiltonian that can be diagonalized numerically.

We start by comparing the topological phase diagrams for Ge channels of different widths. Throughout this entire section, we use $\gamma_1=13.35$, $\gamma_s=4.97$, and $\kappa=3.41$ for the Ge band structure parameters, and the thickness of the well is fixed to $L_z=22$~nm, which is a thickness that is routinely realized in current state-of-the-art experiments~\cite{Scappucci2021}. (Additional phase diagrams for alternative values of $L_z$ are shown in Appendix A.) For now, we further fix the external electric field to $\mathcal{E}_z=0.5$~V$\mu$m$^{-1}$ and we focus on the regime of small strain by choosing $E_s=10$~meV. Assuming that the strain energy depends linearly on the percentage of Si in the barrier and using $E_s=23.7$~meV at $20\%$ (see Ref.~\cite{Sammak2019}) as a reference point, our choice of $E_s=10$~meV corresponds to approximately $8.5\%$ of Si in the barrier, which is only slightly below the Si concentrations of $10\%-20\%$ that are currently used in state-of-the-art devices. The proximity-induced superconducting pairing amplitude is fixed to $\Delta_0=0.1$~meV~\cite{Vigneau2019}. Since the Hamiltonian $H$ belongs to the symmetry class D~\cite{Ryu2010}, the topological transition---if there is any---takes place at $k_x=0$ and is characterized by a change of sign of the $\mathbb{Z}_2$ Pfaffian invariant for 1D topological superconductors~\cite{Kitaev2001,Tewari2012,Budich2013}, which we evaluate numerically. In Figs.~\ref{fig:phase_diagrams_square}(a-d), we show the resulting topological phase diagrams as a function of the magnetic field $B$ and the chemical potential $\mu$ for several channel widths $L_y$ in the case of hard-wall confinement along the $y$ direction. In all cases, the chemical potential is chosen such that only the lowest confinement-induced Ge subband is occupied. The white regions in the phase diagrams correspond to the trivial phase with Pfaffian invariant $+1$, while the colored regions correspond to the topological phase with Pfaffian invariant $-1$, with the color scheme encoding the size of the topological gap (i.e., the bulk gap in the topological phase) obtained by numerical exact diagonalization. We find that the topological phase diagrams resemble the ones that are frequently encountered in the context of standard electron Rashba nanowires~\cite{Oreg2010,Lutchyn2010,Stanescu2011,Sau2012}, which is not very surprising given the form of the effective low-energy Hamiltonian in Eq.~(\ref{eq:eff2band}). However, we stress again that, for hole nanowires, the effective parameters entering Eq.~(\ref{eq:eff2band}) are strongly geometry-dependent. Indeed, we find that the width of the channel has a critical effect on the topological phase diagram, with narrow channels manifesting larger maximal topological gaps [see Fig.~\ref{fig:phase_diagrams_square}(e)] and a significantly larger topological phase space than wide channels, where the topological phase can only be achieved at high magnetic fields close to the critical field of the superconductor.

The strong geometry-dependence of the topological phase diagrams can be understood from the behavior of the effective parameters: First, as the width of the channel increases, the effective $g$ factor $g_\mathrm{eff}$ [see Fig.~\ref{fig:phase_diagrams_square}(f)] decreases significantly due to the decreasing HH-LH mixing. Indeed, it is well known that the effective in-plane $g$ factor of Ge becomes very small as one moves towards the 2D limit where the lowest subband has predominantly HH character~\cite{Scappucci2021,Hendrickx2020,Watzinger2016,Lu2017,Hofmann2019,Gao2020}. Second, the effective spin-orbit energy $E_{so}=\bar{m}\alpha_{so}^2/2\hbar^2$ [see Fig.~\ref{fig:phase_diagrams_square}(g)] reaches a maximum at a relatively small value of $L_y\approx 16$-$17$~nm and decreases significantly as the channel width increases further. It is known from previous works~\cite{Adelsberger2021,Bosco2021} that such a maximum exists and that its exact position and magnitude depend on various system parameters such as the channel geometry, the applied electric field, and strain. Generally, we find that the maximum moves to smaller (larger) channel widths with decreasing (increasing) $L_z$, increasing (decreasing) electric field, and/or increasing (decreasing) strain. As one moves away from the strictly 1D limit with two axes of strongest confinement, the effective SOI becomes very small due to the decreasing HH-LH mixing~\cite{Kloeffel2011,Kloeffel2018}. At this point, it should also be noted that, while the nominal aspect ratio $L_y/L_z$ is not very large even for the widest channels considered here, the electric field induces an additional length scale into the problem, such that the wave function of the lowest-energy subband is compressed along the $z$ direction to a size of $l_\mathcal{E}=(\hbar^2\gamma_1/2me\mathcal{E}_z)^{1/3} \approx 10$~nm for $\mathcal{E}_z=0.5$~V$\mu$m$^{-1}$. We further mention that, since we include orbital effects in our model [see Eq.~(\ref{eq:orb})], both the effective $g$ factor as well as the effective SOI strength $\alpha_{so}$ can in principle depend on the magnetic field. Throughout this paper, we show the effective $g$ factors at $B=2$~T and the effective spin-orbit energies at zero magnetic field. The effective proximity-induced superconducting gap that is opened in the lowest subband of the Ge hole channel remains constant for all of the considered wire geometries, see Fig.~\ref{fig:phase_diagrams_square}(h). We note that this is a direct consequence of our choice of pairing amplitudes $\Delta_{3/2}=-\Delta_{1/2}=\Delta$. Incorporating different pairing amplitudes for HHs and LHs would, within our simplified description of the superconducting pairing given in Eq.~(\ref{eq:sc_pairing}), result in a geometry-dependent renormalization of the effective superconducting gap.

In Figs.~\ref{fig:phase_diagrams_parabolic}(a-d), we show topological phase diagrams for different confinement lengths $l_y$ in the case of parabolic confinement along the $y$ direction. Again, we fix the thickness of the well as $L_z=22$~nm and the strain energy as $E_s=10$~meV, but we choose a larger electric field $\mathcal{E}_z=1$~V$\mu$m$^{-1}$. In general, the achievable topological gaps are smaller than in the case of hard-wall confinement even for narrow channels, see Fig.~\ref{fig:phase_diagrams_parabolic}(e), and larger magnetic fields are required to enter the topological phase. Again, the effective $g$ factor and the effective spin-orbit energy strongly depend on the width of the channel [see Figs.~\ref{fig:phase_diagrams_parabolic}(f) and \ref{fig:phase_diagrams_parabolic}(g)], and both generally take on smaller values than in the case of hard-wall confinement. This can be explained by the reduced level spacing of the subbands induced by the parabolic confinement potential, which leads to a reduced HH-LH mixing. In contrast to the hard-wall case, the SOI decreases monotonically throughout the entire range of confinement lengths considered here. Indeed, Ref.~\cite{Bosco2021} has previously derived the ideal confinement length $l_y$ that maximizes the SOI in a Ge hole channel for a given thickness of the well $L_z$ and a given electric field $\mathcal{E}_z$, finding that, for typical thicknesses $L_z\approx 15$-$30$~nm and an electric field of $\mathcal{E}_z\approx 1$~V$\mu$m$^{-1}$, the ideal confinement length is $l_y\approx 5$-$7$~nm in an unstrained device and even smaller in the presence of strain, which is outside the range of confinement lengths displayed here. As such, if experimentally feasible, the fabrication of extremely narrow channels would further increase the effective $g$ factor and the effective SOI, and, therefore, also the achievable topological gaps. (We note that short ultra-narrow channels with widths of only a few nm have already been fabricated in silicon~\cite{Doris2002}.) Finally, the effective superconducting gap is again independent of the wire geometry, see Fig.~\ref{fig:phase_diagrams_parabolic}(h).

\begin{figure*}[tb]
	\centering
	\includegraphics[width=0.9\textwidth]{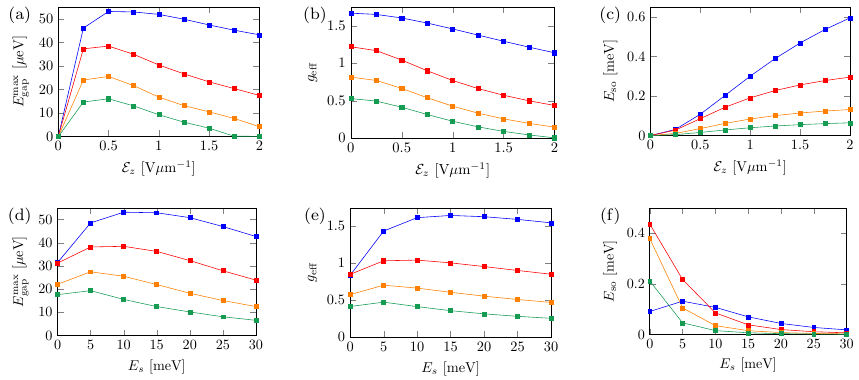}
	\caption{(a,d) Maximal topological gap, (b,e) effective $g$ factor at $B=2$~T, and (c,f) effective spin-orbit energy at zero magnetic field 
	for different channel widths $L_y$ in the case of hard-wall confinement along the $y$ direction (blue: $L_y=15$~nm, red: $L_y=20$~nm, orange: $L_y=25$~nm, green: $L_y=30$~nm). The solid lines are a guide to the eye only. (a-c) Dependence on the external electric field $\mathcal{E}_z$. (d-f) Dependence on the strain energy $E_s$. In all panels, we set $L_z=22$~nm and $\Delta_0=0.1$~meV. In panels (a-c) we fix $E_s=10$~meV and in panels (d-f) we fix $\mathcal{E}_z=0.5$~V$\mu$m$^{-1}$.}
	\label{fig:field_and_strain_dependence_square}
\end{figure*}

Next, we analyze how the maximal topological gap in a given wire geometry depends on the external electric field $\mathcal{E}_z$ and the strain energy $E_s$, both of which have been kept fixed so far. In Fig.~\ref{fig:field_and_strain_dependence_square}(a), we show the maximal topological gap as a function of $\mathcal{E}_z$ for different channel widths $L_y$ in the case of hard-wall confinement along the $y$ direction. We find a non-monotonic dependence that can be explained by the behavior of the effective parameters: On the one hand, the effective $g$ factor generally decreases with increasing electric field, see Fig.~\ref{fig:field_and_strain_dependence_square}(b). This finding is consistent with earlier studies of Ge hole nanowires in the context of spin qubits~\cite{Adelsberger2021,Adelsberger2022,Bosco2021}. On the other hand, the spin-orbit energy grows with the applied electric field throughout the entire range of fields considered here, see Fig.~\ref{fig:field_and_strain_dependence_square}(c). As such, there is a trade-off between a large spin-orbit energy and a large effective $g$ factor, leading, within our model, to a maximal topological gap at a moderate field $\mathcal{E}_z\approx 0.5$~V$\mu$m$^{-1}$. In Fig.~\ref{fig:field_and_strain_dependence_square}(d), we show the maximal topological gap as a function of the strain energy $E_s$. We find that the overall maximum is achieved at moderate strain energies for narrow channels, while the maximum moves to smaller strain energies as the width of the channel increases. This is consistent with the behavior of the effective $g$ factor, see Fig.~\ref{fig:field_and_strain_dependence_square}(e), which shows the same qualitative strain dependence as the maximal topological gap. The spin-orbit energy generally decreases with increasing strain energy, see Fig.~\ref{fig:field_and_strain_dependence_square}(f). 
Finally, we note that, within our simple model, the effective superconducting gap that is opened in the lowest subband of the Ge hole channel is independent of both strain and electric field. However, if the superconducting proximity effect is treated in a more elaborate way that explicitly takes into account the tunneling between the Ge and the superconductor, a field dependence of the tunneling amplitudes and therefore of the effective proximity-induced superconducting gap can be expected since the electric field is responsible for pushing the wave function towards the Ge/superconductor interface~\cite{Adelsberger2023}.

\begin{figure*}[tb]
	\centering
	\includegraphics[width=0.9\textwidth]{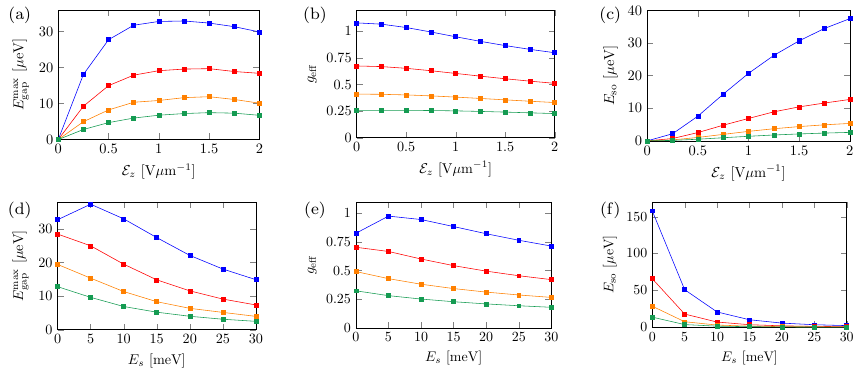}
	\caption{(a,d) Maximal topological gap, (b,e) effective $g$ factor at $B=2$~T, and (c,f) effective spin-orbit energy at zero magnetic field for different confinement lengths $l_y$ in the case of parabolic confinement along the $y$ direction (blue: $l_y=7.5$~nm, red: $l_y=10$~nm, orange: $l_y=12.5$~nm, green: $l_y=15$~nm). The solid lines are a guide to the eye only. (a-c) Dependence on the external electric field $\mathcal{E}_z$. (d-f) Dependence on the strain energy $E_s$. In all panels, we set $L_z=22$~nm and $\Delta_0=0.1$~meV. In panels (a-c) we fix $E_s=10$~meV and in panels (d-f) we fix  $\mathcal{E}_z=1$~V$\mu$m$^{-1}$.}
	\label{fig:field_and_strain_dependence_parabolic}
\end{figure*}

Figure~\ref{fig:field_and_strain_dependence_parabolic} shows the same quantities as Fig.~\ref{fig:field_and_strain_dependence_square} but for the case of a parabolic confinement potential along the $y$ direction. In all panels, the overall trends are consistent with the ones observed for hard-wall confinement, showing that the details of the confinement potential only lead to quantitative, but not qualitative, changes in the behavior of the system. From Fig.~\ref{fig:field_and_strain_dependence_parabolic}(a), we see that the maximal topological gaps are generally smaller than in the hard-wall case, mainly because the parabolic confinement is softer than the hard-wall confinement. Additionally, the maximal topological gaps are shifted to larger electric fields since (1) the dependence of the effective $g$ factor on the electric field is less pronounced [see Fig.~\ref{fig:field_and_strain_dependence_parabolic}(b)] and (2) the spin-orbit energies are significantly reduced compared to the hard-wall case [see Fig.~\ref{fig:field_and_strain_dependence_parabolic}(c)]. As such, the increase in spin-orbit energy with the applied electric field outweighs the detrimental effects of a slightly reduced $g$ factor up to relatively large electric fields, moving the maximal topological gap to $\mathcal{E}_z\approx 1.5$~V$\mu$m$^{-1}$. Figure~\ref{fig:field_and_strain_dependence_parabolic}(d) shows that large values of strain generally reduce the topological gap. In fact, for all but the most narrow channels, the maximal topological gap occurs at zero strain. This is consistent with the behavior of the effective $g$ factor [see Fig.~\ref{fig:field_and_strain_dependence_parabolic}(e)] and the spin-orbit energy [see Fig.~\ref{fig:field_and_strain_dependence_parabolic}(f)], where especially the latter decreases significantly with increasing strain.

In summary, Figs.~\ref{fig:phase_diagrams_square}-\ref{fig:field_and_strain_dependence_parabolic} show that narrow gate-defined channels in lightly strained Ge 2DHGs are a promising platform for the realization of MZMs. On the other hand, wider channels exhibit a small topological phase space that is only accessible at relatively high magnetic fields close to the critical field of Al, making the realization of MZMs challenging. While the effective spin-orbit energy increases with increasing electric field, the effective $g$ factor typically decreases, leading to an optimal regime that can be accessed by tuning the electric field. Within the range of channel widths considered here, the maximal topological gaps are estimated to be on the order of a few tens of $\mu$eV, 
which is comparable to what is expected and observed in InAs systems~\cite{Aghaee2022}. We note that, while we have focused on a Ge/Al heterostructure in this work for concreteness, using a superconductor with a larger critical field (e.g., Nb) would significantly increase the topological phase space and the maximal topological gaps, making the topological phase potentially accessible even in wider channels.

We conclude this section by briefly commenting on several limitations of our model. First, our simulations are based on an effective 4-band model for the top-most valence bands in Ge, while we have neglected the spin split-off band due to its large separation of $\Delta_{SO}\approx 300$~meV. Including the spin split-off band into our description will result in quantitative corrections to our results that become more pronounced as the width of the wire decreases~\cite{Adelsberger2022}. Second, we have used the isotropic approximation of the LK Hamiltonian. If anisotropies are taken into account, the effective parameters such as the effective $g$ factor and the effective SOI become dependent on the growth direction of the quantum well and the orientation of the channel with respect to the crystallographic axes. We expect that the resulting corrections can either reduce or enhance the effective $g$ factor and SOI~\cite{Adelsberger2022} and, therefore, also the maximal topological gaps, presenting an opportunity for further optimization of the device geometry. Third, we have neglected the finite depth of the quantum well. Last but not least, we note that the SOI in wide channels may be underestimated in our description. While our model takes into account the so-called direct SOI that results directly from the 4-band LK Hamiltonian, we have neglected additional contributions to the SOI resulting from couplings to remote bands~\cite{Winkler2003,Kloeffel2011,Kloeffel2018,Gao2020} and interface effects~\cite{Durnev2014,Xiong2021}. While these additional contributions are expected to be negligible in narrow wires, where the direct SOI is very large, they may become significant in wider wires.

\section{Disorder}

It has been now clear for more than 5 years, after the initial short-lived euphoria of the zero-bias tunnel conductance peak observations in InSb- and InAs-based Majorana nanowire platforms~\cite{Mourik2012,Rohkinson2012,Das2012,Deng2012,Lee2012,Churchill2013,Deng2016,Deng2018}, that the current generation of mostly InAs/Al-based semiconductor-superconductor platforms are simply too dirty for the manifestation of non-Abelian MZMs and topological superconductivity because the existing disorder suppresses topology~\cite{Sau2012,Sau2013,Chiu2017,Pan2020,Pan2020b,Dassarma2021,Dassarma2023,Ahn2021,Dassarma2023b,Dassarma2023c}. This is true not just for the early experiments, but also for the latest impressive Microsoft experiment using state-of-the-art InAs samples, where small and fragile topological gaps ($\sim 25~\mu$eV) over small regions of magnetic field and gate voltage were reported very recently~\cite{Aghaee2022}. Recent in-depth independent analyses of the Microsoft data point to the presence of substantial disorder in the system, calling into question whether the observed topological gap and the associated zero modes are generically topological or finite-size mesoscopic fluctuations~\cite{Dassarma2023b,Dassarma2023c}. The estimated disorder in this state-of-the-art InAs platform is of the order of $0.6-1.2$~meV, which is an order of magnitude larger than the claimed topological gap in the Microsoft experiment.  We emphasize that this Microsoft experiment is by far the best measurement in the Majorana nanowire literature with all the earlier nanowire experiments having another order of magnitude larger disorder~\cite{Pan2020,Pan2020b,Dassarma2021,Dassarma2023,Ahn2021}.

A question, therefore, naturally arises why there should be any interest at all in the Ge nanowire platform where the disorder-free pristine topological gap (according to the current calculations presented in this paper) is at best $50~\mu$eV. For a comparison, the corresponding pristine gap is $150-200~\mu$eV in the InAs/Al nanowires without disorder effects. The answer to this question is the extraordinary material quality of the Ge system recently developed in Delft~\cite{Lodari2022,Stehouwer2023}. In fact, our theoretical work is motivated entirely by the extremely high quality of the Ge hole systems developed in Delft.

Using a direct comparison, the best Ge holes and InAs electrons have low-temperature mobilities of $1.2\times10^6 \mathrm{cm}^2/$Vs and $5\times10^4 \mathrm{cm}^2/$Vs, respectively. Using the known effective masses of $0.07m$ (for Ge holes) and $0.02m$ (for InAs electrons), these mobilities can be converted into effective disorder strengths of $\sim 5~\mu$eV (for Ge holes) and $\sim 600~\mu$eV (for InAs electrons).  Note that this estimate ($\sim 0.6$~meV) of the InAs disorder is consistent with Refs.~\cite{Pan2020,Pan2020b,Dassarma2021,Dassarma2023,Ahn2021,Dassarma2023b,Dassarma2023c}, and is in fact a lower bound on the InAs disorder. (This much better quality of the Ge system compared with the InAs system is also reflected in the Ge system having a much lower percolation metal-insulator transition than InAs.)  We are therefore faced with two very contrasting situations:~(1) Electrons in InAs/Al nanowires have a pristine gap $\sim 0.2$~meV and a disorder of $> 0.6$~meV;~(2) holes in Ge/Al nanowires have a pristine gap of $\sim 0.05$~meV and a disorder of $\sim 0.005$~meV. It is clear that this comparison favors the Ge system since the pristine gap, although it is smaller than in InAs, is 10 times the disorder strength whereas in the InAs/Al system, as has already been emphasized in Refs.~\cite{Pan2020,Pan2020b,Dassarma2021,Dassarma2023,Ahn2021,Dassarma2023b,Dassarma2023c}, the disorder is at least 3 times larger than the pristine gap. Earlier works show that the topology perhaps survives a disorder twice the pristine gap, but this constraint has hardly been satisfied in InAs, whereas in Ge, our current work shows that the pristine topological gap is an order of magnitude larger than the low disorder level already achieved in the existing materials. Obviously, a better solution is making the InAs system cleaner, reducing its disorder, but until that happens, Ge/Al is clearly a more promising Majorana platform because it has a much larger gap-to-disorder ratio than InAs. We mention as an aside that InSb nanowires are far worse than InAs nanowires with much larger intrinsic disorder, which is why Microsoft and most other experimental groups have discarded the InSb platform completely.

\section{Conclusions}

We have shown that gate-defined 1D channels in Ge 2DHGs can enter a topological superconducting phase with MZMs at the ends of the channel. We find that the topological gaps are largest (on the order of tens of $\mu$eV) for narrow channels due to strong HH-LH mixing, while both the maximal topological gaps as well as the overall topological phase space are significantly reduced in wider channels due to the small in-plane $g$ factor of the lowest HH-like subband in planar Ge. Large values of strain generally reduce the topological gaps, with the detrimental effect becoming more pronounced as the width of the channel increases and/or the confinement becomes softer. Furthermore, for all of the considered wire geometries, the external electric field provides a tuning knob that can be adjusted in order to maximize the topological gap for a given geometry.

The main advantage of using Ge as a platform for MZMs is the high quality of the material. Since ultra-high quality Ge 2DHGs with very high hole mobilities already exist~\cite{Sammak2019,Lodari2022,Myronov2023,Stehouwer2023}, it is reasonable to expect that high-quality gate-defined Ge hole channels are within experimental reach as well. While the Ge/superconductor interface may introduce additional disorder into the system, first proof-of-principle experiments show that a hard superconducting gap in Ge-based hybrid devices can be achieved, and further experimental progress in this direction is to be expected. As such, despite the relatively small pristine topological gaps found in this work, Ge/superconductor hybrids can potentially exhibit reduced disorder-to-gap ratios compared to hybrid devices based on InAs or InSb, where disorder is likely the most challenging obstacle for future progress. Experimentally, the signatures of MZMs in Ge hole nanowires remain the same as in the InAs or InSb platform (i.e., the main MZM signature is a zero-bias peak in the local conductance), but the reduced disorder-to-gap ratio should lead to a significant reduction of spurious signals stemming from disorder-induced in-gap Andreev bound states, and, therefore, to less ambiguity in the experimental transport data. A detailed analysis of disorder in Ge hole nanowires and its effects on the experimental MZM signatures will be presented in future work. In the context of MZM detection, we further mention that also quasi-Majorana bound states originating from smooth parameter variations and the presence of unintentional quantum dots at the ends of the Ge wire can mimic the signatures of MZMs. We leave an analysis of quasi-MZMs to future work as well.

Our work shows that Ge-based MZM nanowires have serious advantages if very narrow clean Ge channels can be fabricated, which would enhance both the spin-orbit coupling and the $g$ factor, thus enabling topological gaps approaching 50~$\mu$eV in Ge/Al hybrid structures. However, any topological gap larger than 50~$\mu$eV may necessitate using a parent superconductor (e.g. Pb, Nb) with larger gap (and/or larger critical field). An additional considerable advantage of Ge systems is that Ge hole spin qubits can in principle be fabricated on the same Ge device containing MZMs, thus enabling a combination of circuit level and topological quantum computation in a monolithic structure.

\section*{Acknowledgments} 
We thank A. R. Akhmerov for helpful discussions on the form of the superconducting pairing. This work is supported by the Laboratory for Physical Sciences through the Condensed Matter Theory Center.

\appendix

\section{Additional phase diagrams}

\begin{figure*}[tb]
	\centering
	\includegraphics[width=\textwidth]{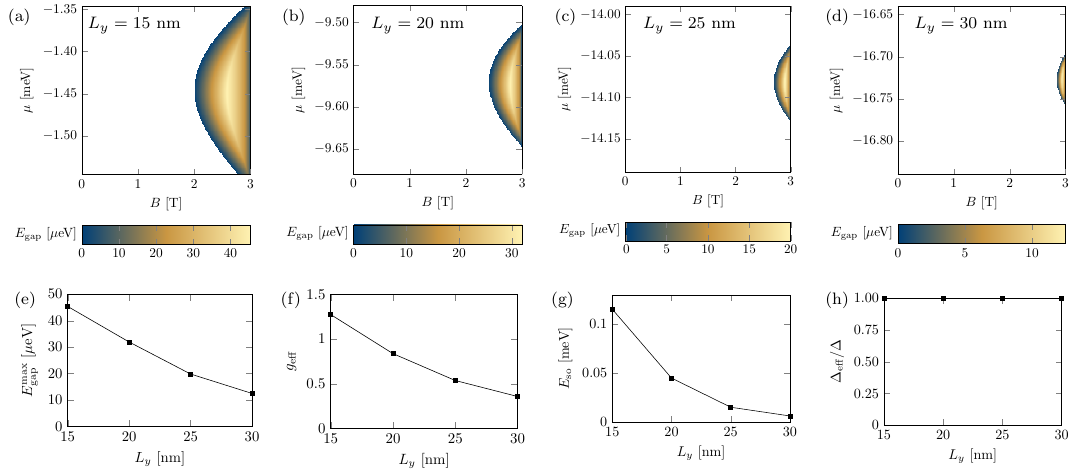}
	\caption{(a-d) Topological phase diagrams obtained by numerically diagonalizing $H=H_0+H_{sc}$ for different widths of the channel $L_y$ (see insets) in the case of hard-wall confinement along the $y$ direction. The white regions correspond to the trivial phase, while the colored regions correspond to the topological phase with the color encoding the size of the bulk gap. (e) Maximal bulk gap in the topological phase in dependence on the channel width. (f-h) Effective $g$ factor at $B=2$~T, effective spin-orbit energy at zero magnetic field, and effective superconducting gap at zero magnetic field in dependence on the channel width. The solid lines are a guide to the eye only. We fix $L_z=18$~nm, $\Delta_0=0.1$~meV, $E_s=10$~meV, and $\mathcal{E}_z=0.5$~V$\mu$m$^{-1}$ for all panels.}
	\label{fig:phase_diagrams_Lz_18}
\end{figure*}

\begin{figure*}[tb]
	\centering
	\includegraphics[width=0.9\textwidth]{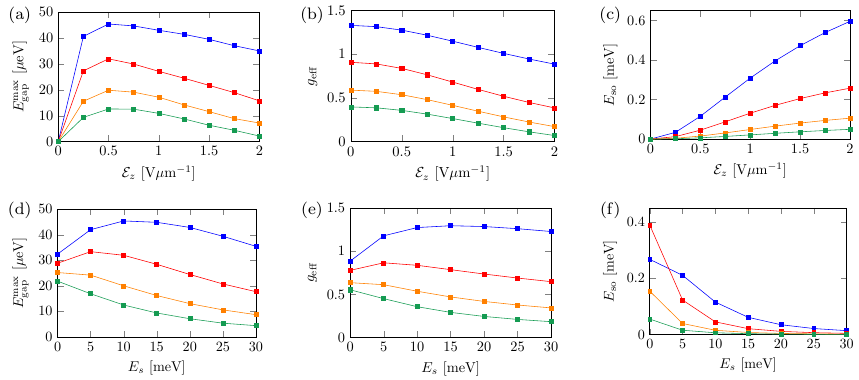}
	\caption{(a,d) Maximal topological gap, (b,e) effective $g$ factor at $B=2$~T, and (c,f) effective spin-orbit energy at zero magnetic field 
		for different channel widths $L_y$ in the case of hard-wall confinement along the $y$ direction (blue: $L_y=15$~nm, red: $L_y=20$~nm, orange: $L_y=25$~nm, green: $L_y=30$~nm). The solid lines are a guide to the eye only. (a-c) Dependence on the external electric field $\mathcal{E}_z$. (d-f) Dependence on the strain energy $E_s$. In all panels, we set $L_z=18$~nm and $\Delta_0=0.1$~meV. In panels (a-c) we fix $E_s=10$~meV and in panels (d-f) we fix $\mathcal{E}_z=0.5$~V$\mu$m$^{-1}$.}
	\label{fig:field_and_strain_dependence_Lz_18}
\end{figure*}

\begin{figure*}[tb]
	\centering
	\includegraphics[width=\textwidth]{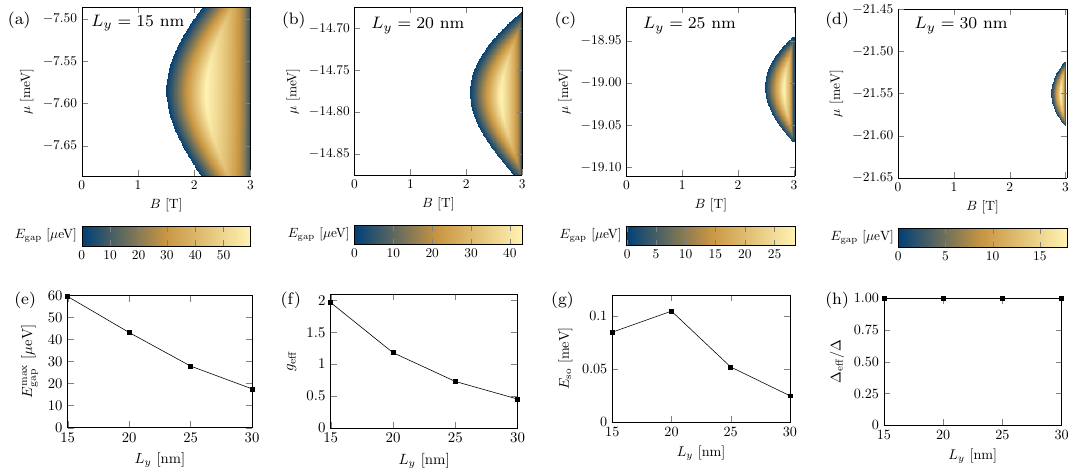}
	\caption{(a-d) Topological phase diagrams obtained by numerically diagonalizing $H=H_0+H_{sc}$ for different widths of the channel $L_y$ (see insets) in the case of hard-wall confinement along the $y$ direction. The white regions correspond to the trivial phase, while the colored regions correspond to the topological phase with the color encoding the size of the bulk gap. (e) Maximal bulk gap in the topological phase in dependence on the channel width. (f-h) Effective $g$ factor at $B=2$~T, effective spin-orbit energy at zero magnetic field, and effective superconducting gap at zero magnetic field in dependence on the channel width. The solid lines are a guide to the eye only. We fix $L_z=26$~nm, $\Delta_0=0.1$~meV, $E_s=10$~meV, and $\mathcal{E}_z=0.5$~V$\mu$m$^{-1}$ for all panels.}
	\label{fig:phase_diagrams_Lz_26}
\end{figure*}

\begin{figure*}[tb]
	\centering
	\includegraphics[width=0.9\textwidth]{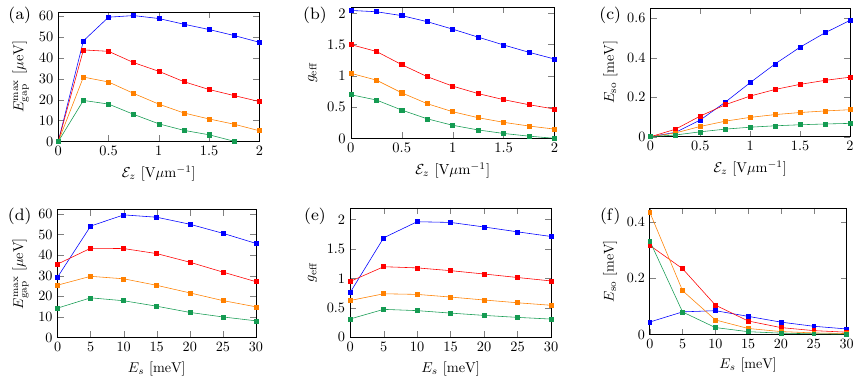}
	\caption{(a,d) Maximal topological gap, (b,e) effective $g$ factor at $B=2$~T, and (c,f) effective spin-orbit energy at zero magnetic field 
		for different channel widths $L_y$ in the case of hard-wall confinement along the $y$ direction (blue: $L_y=15$~nm, red: $L_y=20$~nm, orange: $L_y=25$~nm, green: $L_y=30$~nm). The solid lines are a guide to the eye only. (a-c) Dependence on the external electric field $\mathcal{E}_z$. (d-f) Dependence on the strain energy $E_s$. In all panels, we set $L_z=26$~nm and $\Delta_0=0.1$~meV. In panels (a-c) we fix $E_s=10$~meV and in panels (d-f) we fix $\mathcal{E}_z=0.5$~V$\mu$m$^{-1}$.}
	\label{fig:field_and_strain_dependence_Lz_26}
\end{figure*}

Throughout the main text, the thickness of the quantum well has been kept fixed to $L_z=22$~nm. In this appendix, we present numerical results for two additional thicknesses $L_z=18$~nm and $L_z=26$~nm. In Fig.~\ref{fig:phase_diagrams_Lz_18} (Fig.~\ref{fig:phase_diagrams_Lz_26}), we show topological phase diagrams for $L_z=18$~nm ($L_z=26$~nm) for the case of hard-wall confinement along the $y$ direction for different widths $L_y$. In Fig.~\ref{fig:field_and_strain_dependence_Lz_18} (Fig.~\ref{fig:field_and_strain_dependence_Lz_26}), we show the corresponding maximal topological gaps as a function of electric field and strain. For both thicknesses, we find that the general trends discussed in the main text persist. In particular, the topological gaps, the effective $g$ factors, and the effective SOI generally decrease as the width of the channel increases. While the $g$ factor decreases monotonically throughout the range of channel widths considered here, the SOI exhibits a maximum at a small value of $L_y$ that depends on the thickness of the quantum well $L_z$, the electric length $l_\mathcal{E}$, and the strain energy $E_s$. In Fig.~\ref{fig:field_and_strain_dependence_Lz_18}(c), this maximum occurs outside the range of widths considered here, while it can be seen to occur around $L_y\approx 20$~nm in Fig.~\ref{fig:field_and_strain_dependence_Lz_26}(c). Furthermore, the topological gap exhibits a maximum in dependence on the electric field due to a trade-off between an increasing (decreasing) effective spin-orbit energy (effective $g$ factor) as the electric field is increased. Large values of strain generally decrease the topological gaps except for very narrow wires with strong HH-LH mixing.
	
\begin{figure*}[tb]
	\centering
	\includegraphics[width=\textwidth]{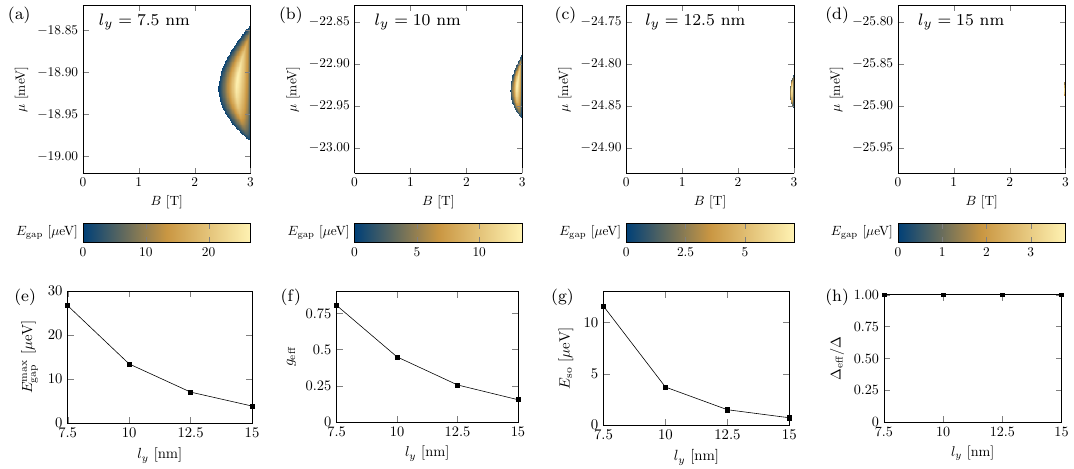}
\caption{(a-d) Topological phase diagrams obtained by numerically diagonalizing $H=H_0+H_{sc}$ for different confinement lengths $l_y$ (see insets) in the case of parabolic confinement along the $y$ direction. The white regions correspond to the trivial phase, while the colored regions correspond to the topological phase with the color encoding the size of the bulk gap. (e) Maximal bulk gap in the topological phase in dependence on the confinement length. (f-h) Effective $g$ factor at $B=2$~T, effective spin-orbit energy at zero magnetic field, and effective superconducting gap at zero magnetic field in dependence on the confinement length. The solid lines are a guide to the eye only. We fix $L_z=18$~nm, $\Delta_0=0.1$~meV, $E_s=10$~meV, and $\mathcal{E}_z=1$~V$\mu$m$^{-1}$ for all panels.}
	\label{fig:phase_diagrams_parabolic_Lz_18}
\end{figure*}

\begin{figure*}[tb]
	\centering
	\includegraphics[width=0.9\textwidth]{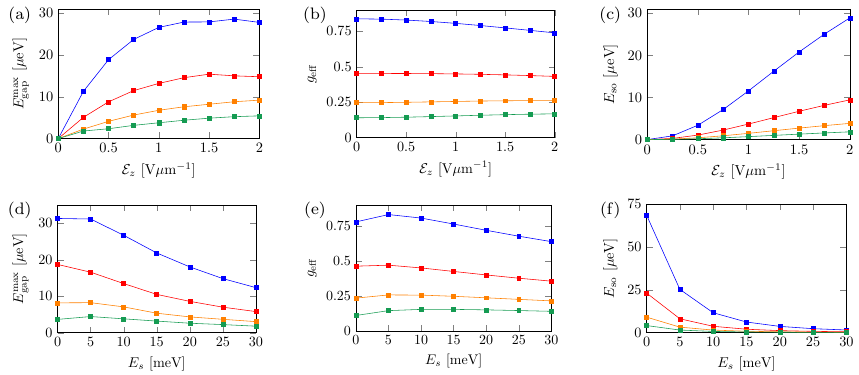}
	\caption{(a,d) Maximal topological gap, (b,e) effective $g$ factor at $B=2$~T, and (c,f) effective spin-orbit energy at zero magnetic field 
		for different confinement lengths $l_y$ in the case of parabolic confinement along the $y$ direction (blue: $l_y=7.5$~nm, red: $l_y=10$~nm, orange: $l_y=12.5$~nm, green: $l_y=15$~nm). The solid lines are a guide to the eye only. (a-c) Dependence on the external electric field $\mathcal{E}_z$. (d-f) Dependence on the strain energy $E_s$. In all panels, we set $L_z=18$~nm and $\Delta_0=0.1$~meV. In panels (a-c) we fix $E_s=10$~meV and in panels (d-f) we fix  $\mathcal{E}_z=1$~V$\mu$m$^{-1}$.}
	\label{fig:field_and_strain_dependence_parabolic_Lz_18}
\end{figure*}

\begin{figure*}[tb]
	\centering
	\includegraphics[width=\textwidth]{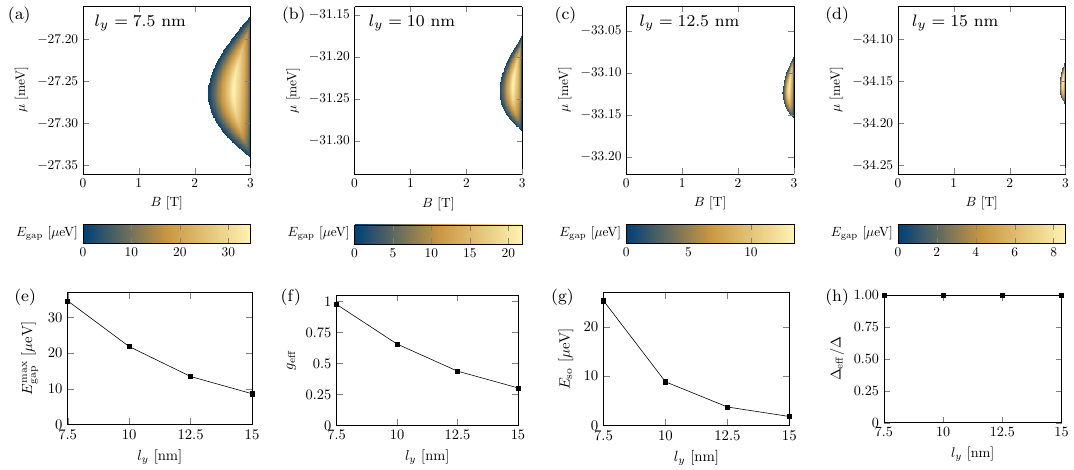}
	\caption{(a-d) Topological phase diagrams obtained by numerically diagonalizing $H=H_0+H_{sc}$ for different confinement lengths $l_y$ (see insets) in the case of parabolic confinement along the $y$ direction. The white regions correspond to the trivial phase, while the colored regions correspond to the topological phase with the color encoding the size of the bulk gap. (e) Maximal bulk gap in the topological phase in dependence on the confinement length. (f-h) Effective $g$ factor at $B=2$~T, effective spin-orbit energy at zero magnetic field, and effective superconducting gap at zero magnetic field in dependence on the confinement length. The solid lines are a guide to the eye only. We fix $L_z=26$~nm, $\Delta_0=0.1$~meV, $E_s=10$~meV, and $\mathcal{E}_z=1$~V$\mu$m$^{-1}$ for all panels.}
	\label{fig:phase_diagrams_parabolic_Lz_26}
\end{figure*}

\begin{figure*}[tb]
	\centering
	\includegraphics[width=0.9\textwidth]{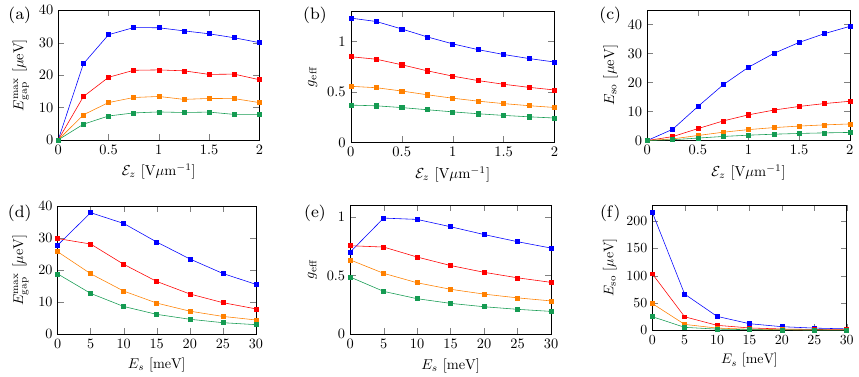}
	\caption{(a,d) Maximal topological gap, (b,e) effective $g$ factor at $B=2$~T, and (c,f) effective spin-orbit energy at zero magnetic field 
		for different confinement lengths $l_y$ in the case of parabolic confinement along the $y$ direction (blue: $l_y=7.5$~nm, red: $l_y=10$~nm, orange: $l_y=12.5$~nm, green: $l_y=15$~nm). The solid lines are a guide to the eye only. (a-c) Dependence on the external electric field $\mathcal{E}_z$. (d-f) Dependence on the strain energy $E_s$. In all panels, we set $L_z=26$~nm and $\Delta_0=0.1$~meV. In panels (a-c) we fix $E_s=10$~meV and in panels (d-f) we fix  $\mathcal{E}_z=1$~V$\mu$m$^{-1}$.}
	\label{fig:field_and_strain_dependence_parabolic_Lz_26}
\end{figure*}

Fig.~\ref{fig:phase_diagrams_parabolic_Lz_18} (Fig.~\ref{fig:phase_diagrams_parabolic_Lz_26}) shows phase diagrams for $L_z=18$~nm ($L_z=26$~nm) for the case of parabolic confinement along the $y$ direction with different confinement lengths $l_y$, and Fig.~\ref{fig:field_and_strain_dependence_parabolic_Lz_18} (Fig.~\ref{fig:field_and_strain_dependence_parabolic_Lz_26}) shows the corresponding maximal topological gaps as a function of electric field and strain. Again, we observe similar trends as in the main text, with the topological gaps, the effective $g$ factors, and the effective SOI decreasing monotonically with $l_y$ throughout the entire range of confinement lengths considered here. There is once again a trade-off between an increasing (decreasing) effective spin-orbit energy (effective $g$ factor) as the electric field is increased, but it is interesting to note that the dependence of the effective $g$ factor on the electric field  [see Figs.~11(b) and 13(b)] is less pronounced than in the case of hard-wall confinement (in fact, we even observe a slight increase of the effective $g$ factor with increasing electric field at $L_z=18$~nm for $l_y=12.5,15$~nm), such that the maximal topological gap moves to larger electric fields compared to the hard-wall case.

\clearpage
\begin{thebibliography}{}
	
	
\bibitem{Scappucci2021}
G. Scappucci, C. Kloeffel, F. A. Zwanenburg, D. Loss, M. Myronov, J.-J. Zhang, S. De Franceschi, G. Katsaros, and M. Veldhorst, Nat. Rev. Mater. {\bf 6}, 926 (2021).


\bibitem{Hu2007}
Y. Hu, H. O. Churchill, D. J. Reilly, J. Xiang, C. M. Lieber, and C. M. Marcus, Nature Nanotech. {\bf 2}, 622 (2007). 
\bibitem{Hu2012}
Y. Hu, F. Kuemmeth, C. M. Lieber, and C. M. Marcus, Nature Nanotech. {\bf 7}, 47 (2012). 
\bibitem{Ares2013}
N. Ares, G. Katsaros, V. N. Golovach, J. Zhang, A. Prager, L. I. Glazman, O. G. Schmidt, and S. De Franceschi, Appl. Phys. Lett. {\bf 103}, 263113 (2013). 
\bibitem{Watzinger2018}
H. Watzinger, J. Kuku\v{c}ka, L. Vuku\v{s}i\'{c}, F. Gao, T. Wang, F. Sch\"{a}ffler, J.-J. Zhang, and G. Katsaros, Nature Comm. {\bf 9}, 3902 (2018). 
\bibitem{Li2018}
Y. Li, S.-X. Li, F. Gao, H.-O. Li, G. Xu, K. Wang, D. Liu, G. Cao, M. Xiao, T. Wang, J.-J. Zhang, G.-C. Guo, and G.-P. Guo, Nano Lett. {\bf 18}, 2091 (2018). 
\bibitem{Hofmann2019}
A. Hofmann, D. Jirovec, M. Borovkov, I. Prieto, A. Ballabio, J. Frigerio, D. Chrastina, G. Isella, and G. Katsaros,
arXiv:1910.05841 (2019).
\bibitem{Hendrickx2020} 
N. W. Hendrickx, W. I. L. Lawrie, L. Petit, A. Sammak, G. Scappucci, and M. Veldhorst, Nat. Commun. {\bf 11}, 3478 (2020).
\bibitem{Jirovec2021}
D. Jirovec, A. Hofmann, A. Ballabio, P. M. Mutter, G. Tavani, M. Botifoll, A. Crippa, J. Kuku\v{c}ka, O. Sagi,
F. Martins, \emph{et al.}, Nat. Mater. {\bf 20}, 1106 (2021). 
\bibitem{Froning2021}
F. N. M. Froning, L. C. Camenzind, O. A. H. van der Molen, A. Li, E. P. A. M. Bakkers, D. M. Zumbühl, and F. R. Braakman, Nature Nanotech. {\bf 16}, 308 (2021).
\bibitem{Wang2022}
K. Wang, G. Xu, F. Gao, H. Liu, R.-L. Ma, X. Zhang, Z. Wang, G. Cao, T. Wang, J.-J. Zhang, D. Culcer, X. Hu, H.-W. Jiang, H.-O. Li, G.-C. Guo, and G.-P. Guo, Nat. Comm. {\bf 13}, 206 (2022). 


\bibitem{Sammak2019}
A. Sammak, D. Sabbagh, N. W. Hendrickx, M. Lodari, B. Paquelet Wuetz, A. Tosato, L. Yeoh, M. Bollani, M. Virgilio, M. A. Schubert, P. Zaumseil, G. Capellini, M. Veldhorst, and G. Scappucci, Adv. Funct. Mater. {\bf 29}, 1807613 (2019).
\bibitem{Lodari2022}
M. Lodari, O. Kong, M. Rendell, A. Tosato, A. Sammak, M. Veldhorst, A. R. Hamilton, and G. Scappucci, Appl. Phys. Lett. {\bf 120}, 122104 (2022).
\bibitem{Myronov2023}
M. Myronov, J. Kycia, P. Waldron, W. Jiang, P. Barrios, A. Bogan, P. Coleridge, and S. Studenikin, Small
Science {\bf 3}, 2200094 (2023).
\bibitem{Stehouwer2023}
L. E. A. Stehouwer, A. Tosato, D. Degli Esposti, D. Costa, M. Veldhorst, A. Sammak, and G. Scappucci, Appl. Phys. Lett. {\bf 123}, 092101 (2023).


\bibitem{Lawrie2020}
W. I. L. Lawrie, H. G. J. Eenink, N. W. Hendrickx, J. M. Boter, L. Petit, S. V. Amitonov, M. Lodari, B. Paquelet Wuetz, C. Volk, S. G. J. Philips, G. Droulers, N. Kalhor, F. van Riggelen, D. Brousse, A. Sammak, L. M. K. Vandersypen, G. Scappucci, and M. Veldhorst, Appl. Phys. Lett. {\bf 116}, 080501 (2020).
\bibitem{Riggelen2020}
F. van Riggelen, N. W. Hendrickx, W. I. L. Lawrie, M. Russ, A. Sammak, G. Scappucci, and M. Veldhorst, Appl. Phys. Lett. {\bf 118}, 044002 (2021).
\bibitem{Hendrickx2021}
N. W. Hendrickx, W. I. Lawrie, M. Russ, F. van Riggelen, S. L. de Snoo, R. N. Schouten, A. Sammak, G. Scappucci,
and M. Veldhorst, Nature {\bf 591}, 580 (2021). 


\bibitem{Hendrickx2020b}
N. Hendrickx, D. Franke, A. Sammak, G. Scappucci, and M. Veldhorst, Nature {\bf 577}, 487 (2020). 



\bibitem{Vries2018}
F. K. de Vries, J. Shen, R. J. Skolasinski, M. P. Nowak, D. Varjas, L. Wang, M. Wimmer, J. Ridderbos, F. A. Zwanenburg, A. Li, S. Koelling, M. A. Verheijen, E. P. A. M. Bakkers, and L. P. Kouwenhoven, Nano Lett. {\bf 18}, 6483 (2018).
\bibitem{Hendrickx2018}
N. Hendrickx, D. Franke, A. Sammak, M. Kouwenhoven, D. Sabbagh, L. Yeoh, R. Li, M. Tagliaferri, M. Virgilio, G. Capellini, \emph{et al.}, Nat. Comm. {\bf 9}, 2835 (2018).
\bibitem{Hendrickx2019}
N. W. Hendrickx, M. L. V. Tagliaferri, M. Kouwenhoven, R. Li, D. P. Franke, A. Sammak, A. Brinkman, G. Scappucci, and M. Veldhorst, Phys. Rev. B {\bf 99}, 075435 (2019).
\bibitem{Vigneau2019}
F. Vigneau, R. Mizokuchi, D. C. Zanuz, X. Huang, S. Tan, R. Maurand, S. Frolov, A. Sammak, G. Scappucci, F. Lefloch, and S. De Franceschi, Nano Lett. {\bf 19}, 1023 (2019).
\bibitem{Aggarwal2021}
K. Aggarwal, A. Hofmann, D. Jirovec, I. Prieto, A. Sammak, M. Botifoll, S. Mart\'{i}-S\'{a}nchez, M. Veldhorst, J. Arbiol, G. Scappucci, J. Danon, and G. Katsaros, Phys. Rev. Res. {\bf 3}, L022005 (2021).
\bibitem{Tosato2022}
A. Tosato, V. Levajac, J.-Y. Wang, C. J. Boor, F. Borsoi, M. Botifoll, C. N. Borja, S. Mart\'{i}-S\'{a}nchez, J. Arbiol,
A. Sammak, M. Veldhorst, and G. Scappucci, Commun. Mater. {\bf 4}, 23 (2023).


\bibitem{Mourik2012}
V. Mourik, K. Zuo, S. M. Frolov, S. R. Plissard, E. P. A. M. Bakkers, and L. P. Kouwenhoven, Science {\bf 336}, 6084 (2012).
\bibitem{Rohkinson2012}
L. P. Rokhinson, X. Liu, and J. K. Furdyna, Nat. Phys. {\bf 8}, 795 (2012).
\bibitem{Das2012}
A. Das, Y. Ronen, Y. Most, Y. Oreg, M. Heiblum, and H. Shtrikman, Nat. Phys. {\bf 8}, 887 (2012).
\bibitem{Deng2012}
M. T. Deng, C. L. Yu, G. Y. Huang, M. Larsson, P. Caroff, and H. Q. Xu, Nano Lett. {\bf 12}, 6414 (2012).
\bibitem{Lee2012}
E. J. H. Lee, X. Jiang, R. Aguado, G. Katsaros, C. M. Lieber, and S. De Franceschi, Phys. Rev. Lett. {\bf 109},
186802 (2012).
\bibitem{Churchill2013}
H. O. H. Churchill, V. Fatemi, K. Grove-Rasmussen, M. T. Deng, P. Caroff, H. Q. Xu, and C. M. Marcus, Phys. Rev. B {\bf 87}, 241401(R) (2013).
\bibitem{Deng2016}
M. T. Deng, S. Vaitiekenas, E. B. Hansen, J. Danon, M. Leijnse, K. Flensberg, J. Nyg\r{a}rd, P. Krogstrup, and C. M. Marcus, Science {\bf 354}, 1557 (2016).
\bibitem{Deng2018}
M. T. Deng, S. Vaitiekenas, E. Prada, P. San-Jose, J. Nyg\r{a}rd, P. Krogstrup, R. Aguado, and C. M. Marcus, Phys. Rev. B {\bf 98}, 085125 (2018).


\bibitem{Sau2012}
J. D. Sau, S. Tewari, and S. Das Sarma, Phys. Rev. B {\bf 85}, 064512 (2012).
\bibitem{Sau2013}
J. D. Sau and S. Das Sarma, Phys. Rev. B {\bf 88}, 064506 (2013).
\bibitem{Chiu2017}
C.-K. Chiu, J. D. Sau, and S. Das Sarma, Phys. Rev. B {\bf 96}, 054504 (2017).
\bibitem{Pan2020}
H. Pan and S. Das Sarma, Phys. Rev. Research {\bf 2}, 013377 (2020).
\bibitem{Pan2020b}
H. Pan, W. S. Cole, J. D. Sau, and S. Das Sarma, Phys. Rev. B {\bf 101}, 024506 (2020).
\bibitem{Dassarma2021}
S. Das Sarma and H. Pan, Phys. Rev. B {\bf 103}, 195158 (2021).
\bibitem{Dassarma2023}
S. Das Sarma, Nat. Phys. {\bf 19}, 165 (2023).
\bibitem{Ahn2021}
S. Ahn, H. Pan, B. Woods, T. D. Stanescu, and S. Das Sarma, Phys. Rev. Materials {\bf 5}, 124602 (2021).
\bibitem{Dassarma2023b}
S. Das Sarma, J. D. Sau, and T. D. Stanescu, Phys. Rev. B {\bf 108}, 085416 (2023).
\bibitem{Dassarma2023c}
S. Das Sarma and H. Pan, Phys. Rev. B {\bf 108}, 085415 (2023).


\bibitem{Aghaee2022}
M. Aghaee \emph{et al.}, Phys. Rev. B {\bf 107}, 245423 (2023).


\bibitem{Kloeffel2011}
C. Kloeffel, M. Trif, and D. Loss, Phys. Rev. B {\bf 84}, 195314 (2011).
\bibitem{Kloeffel2018}
C. Kloeffel, M. J. Ran\v{c}i\'{c}, and D. Loss, Phys. Rev. B {\bf 97}, 235422 (2018).

\bibitem{Maier2014}
F. Maier, J. Klinovaja, and D. Loss, Phys. Rev. B {\bf 90}, 195421 (2014).
\bibitem{Luethi2022}
M. Luethi, K. Laubscher, S. Bosco, D. Loss, and J. Klinovaja, Phys. Rev. B {\bf 107}, 035435 (2023).
\bibitem{Luethi2023}
M. Luethi, H. F. Legg, K. Laubscher, D. Loss, and J. Klinovaja, Phys. Rev. B {\bf 108}, 195406 (2023).


\bibitem{Mao2012}
L. Mao, M. Gong, E. Dumitrescu, S. Tewari, and C. Zhang, Phys. Rev. Lett. {\bf 108}, 177001 (2012).
\bibitem{Liang2017}
J. Liang and Y. Lyanda-Geller, Phys. Rev. B {\bf 95}, 201404(R) (2017).


\bibitem{Mizokuchi2018}
R. Mizokuchi, R. Maurand, F. Vigneau, M. Myronov, and S. De Franceschi, Nano Lett. {\bf 18}, 4861 (2018).


\bibitem{Luttinger1956}
J. M. Luttinger, Phys. Rev. {\bf 102}, 1030 (1956).
\bibitem{Winkler2003}
R. Winkler, {\it Spin-orbit coupling effects in two-dimensional electron and hole systems} (Springer-Verlag, Berlin, Heidelberg, New York, 2003).


\bibitem{Bir1974}
G. L. Bir, G. E. Pikus, \emph{et al.}, {\it Symmetry and strain-induced effects in semiconductors}, Vol. 484 (Wiley, New
York, 1974).


\bibitem{Winkler2000}
R. Winkler, Phys. Rev. B {\bf 62}, 4245 (2000).
\bibitem{Winkler2008}
R. Winkler, D. Culcer, S. J. Papadakis, B. Habib, and M. Shayegan, Semicond. Sci. Technol. {\bf 23}, 114017 (2008).
\bibitem{Moriya2014}
R. Moriya, K. Sawano, Y. Hoshi, S. Masubuchi, Y. Shiraki, A. Wild, C. Neumann, G. Abstreiter, D. Bougeard, T. Koga, and T. Machida, Phys. Rev. Lett. {\bf 113}, 086601 (2014).
\bibitem{Marcellina2017}
E. Marcellina, A. R. Hamilton, R. Winkler, and D. Culcer, Phys. Rev. B {\bf 95}, 075305 (2017).
\bibitem{Mizokuchi2017}
R. Mizokuchi, P. Torresani, R. Maurand, M. Myronov, and S. De Franceschi, Appl. Phys. Lett. {\bf 111}, 063102 (2017).
\bibitem{Chou2018}
C.-T. Chou,   N. T. Jacobson,  J. E. Moussa, A. D. Baczewski, Y. Chuang, C.-Y. Liu, J.-Y. Li, and T. M. Lu,
Nanoscale {\bf 10}, 20559 (2018).
\bibitem{Terrazos2021}
L. A. Terrazos, E. Marcellina, Z. Wang, S. N. Coppersmith, M. Friesen, A. R. Hamilton, X. Hu, B. Koiller, A. L. Saraiva, D. Culcer, and R. B. Capaz, Phys. Rev. B {\bf 103}, 125201 (2021).


\bibitem{Adelsberger2021}
C. Adelsberger, M. Benito, S. Bosco, J. Klinovaja, and D. Loss, Phys. Rev. B {\bf 105}, 075308 (2022).
\bibitem{Adelsberger2022}
C. Adelsberger, S. Bosco, J. Klinovaja, and D. Loss, Phys. Rev. B {\bf 106}, 235408 (2022).

\bibitem{Gao2020}
F. Gao, J.-H. Wang, H. Watzinger, H. Hu, M. J. Ran\v{c}i\'{c}, J.-Y. Zhang, T. Wang, Y. Yao, G.-L. Wang, J. Kuku\v{c}ka, L. Vuku\v{s}i\'{c}, C. Kloeffel, D. Loss, F. Liu, G. Katsaros, and J.-J. Zhang, Adv. Mater. {\bf 32}, 1906523 (2020). 


\bibitem{Hao2010}
X.-J. Hao, T. Tu, G. Cao, C. Zhou, H.-O. Li, G.-C. Guo, W. Y. Fung, Z. Ji, G.-P. Guo, and W. Lu, Nano Lett. {\bf 10}, 2956 (2010).
\bibitem{Froning2021b}
F. N. M. Froning, M. J. Ran\v{c}i\'{c}, B. Het\'{e}nyi, S. Bosco, M. K. Rehmann, A. Li, E. P. A. M. Bakkers, F. A. Zwanenburg, D. Loss, D. M. Zumb\"{u}hl, and F. R. Braakman, Phys. Rev. Res. {\bf 3}, 013081 (2021).


\bibitem{Csontos2009}
D. Csontos, P. Brusheim, U. Z\"{u}licke, and H. Q. Xu, Phys. Rev. B {\bf 79}, 155323 (2009).
\bibitem{Sercel1990}
P. C. Sercel and K. J. Vahala, Phys. Rev. B {\bf 42}, 3690 (1990).


\bibitem{Lawaetz1971}
P. Lawaetz, Phys. Rev. B {\bf 4}, 3460 (1971).

\bibitem{note1}
We note that the opposite choice of sign $\Delta_{3/2}=\Delta_{1/2}\equiv\Delta$ violates a no-go theorem for $s$-wave pairing pointed out in Ref.~\cite{Poyhonen2021}.

\bibitem{Poyhonen2021}
K. P\"{o}yh\"{o}nen, D. Varjas, M. Wimmer, and A. R. Akhmerov, SciPost Phys. {\bf 10}, 108 (2021).

\bibitem{Nichele2017}
F. Nichele, A. C. C. Drachmann, A. M. Whiticar, E. C. T. O'Farrell, H. J. Suominen, A. Fornieri, T. Wang, G. C. Gardner, C. Thomas, A. T. Hatke, P. Krogstrup, M. J. Manfra, K. Flensberg, and C. M. Marcus, Phys. Rev. Lett. {\bf 119}, 136803 (2017).
\bibitem{Suominen2017}
H. J. Suominen, M. Kjaergaard, A. R. Hamilton, J. Shabani, C. J. Palmstr\o{}m, C. M. Marcus, and F. Nichele, Phys. Rev. Lett. {\bf 119}, 176805 (2017).


\bibitem{Sau2010}
J. D. Sau, R. M. Lutchyn, S. Tewari, and S. Das Sarma, Phys. Rev. B {\bf 82}, 094522 (2010).
\bibitem{Stanescu2010}
T. D. Stanescu, J. D. Sau, R. M. Lutchyn, and S. Das Sarma, Phys. Rev. B {\bf 81}, 241310(R) (2010).
\bibitem{Cole2015}
W. S. Cole, S. Das Sarma, and T. D. Stanescu, Phys. Rev. B {\bf 92}, 174511 (2015).
\bibitem{Reeg2018}
C. Reeg, D. Loss, and J. Klinovaja, Phys. Rev. B {\bf 97}, 165425 (2018).
\bibitem{Adelsberger2023}
C. Adelsberger, H. F. Legg, D. Loss, and J. Klinovaja, Phys. Rev. B {\bf 108}, 155433 (2023).

\bibitem{Milivojevic2021}
M. Milivojevi\'{c}, Phys. Rev. B {\bf 104}, 235304 (2021).


\bibitem{Ryu2010}
S. Ryu, A. P. Schnyder, A. Furusaki, and A. W. W. Ludwig, New J. Phys. {\bf 12}, 065010 (2010).


\bibitem{Kitaev2001}
A. Y. Kitaev, Phys. Usp. {\bf 44}, 131 (2001).
\bibitem{Tewari2012}
S. Tewari and J. D. Sau, Phys. Rev. Lett. {\bf 109}, 150408 (2012).
\bibitem{Budich2013}
J. C. Budich and E. Ardonne, Phys. Rev. B {\bf 88}, 075419 (2013).


\bibitem{Lutchyn2010}
R. M. Lutchyn, J. D. Sau, and S. Das Sarma, Phys. Rev. Lett. {\bf 105}, 077001 (2010).
\bibitem{Oreg2010}
Y. Oreg, G. Refael, and F. von Oppen, Phys. Rev. Lett. {\bf 105}, 177002 (2010).
\bibitem{Stanescu2011}
T. D. Stanescu, R. M. Lutchyn, and S. Das Sarma, Phys. Rev. B {\bf 84}, 144522 (2011).


\bibitem{Watzinger2016}
H. Watzinger, C. Kloeffel, L. Vuku\v{s}i\'{c}, M. D. Rossell, V. Sessi, J. Kuku\v{c}ka, R. Kirchschlager, E. Lausecker,
A. Truhlar, M. Glaser, A. Rastelli, A. Fuhrer, D. Loss, and G. Katsaros, Nano Lett. {\bf 16}, 6879 (2016).
\bibitem{Lu2017}
T. Lu, C. Harris, S.-H. Huang, Y. Chuang, J.-Y. Li, and C. Liu, Appl. Phys. Lett. {\bf 111}, 102108 (2017).


\bibitem{Bosco2021}
S. Bosco, M. Benito, C. Adelsberger, and D. Loss, Phys. Rev. B {\bf 104}, 115425 (2021).


\bibitem{Doris2002}
B. Doris, M. Ieong, T. Kanarsky, Y. Zhang, R. A. Roy, O. Dokumaci, Z. Ren, F.-F. Jamin, L. Shi, W. Natzle, H.-J. Huang, J. Mezzapelle, A. Mocuta, S. Womack, M. Gribelyuk, E. C. Jones, R. J. Miller, H.-S. P. Wong, and W. Haensch, Digest. International Electron Devices Meeting, 267 (2002).


\bibitem{Durnev2014}
M. V. Durnev, M. M. Glazov, and E. L. Ivchenko, Phys. Rev. B {\bf 89}, 075430 (2014).
\bibitem{Xiong2021}
J.-X. Xiong, S. Guan, J.-W. Luo, and S.-S. Li, Phys. Rev. B {\bf 103}, 085309 (2021).





	
\end{thebibliography}
\end{document}